\documentclass[11pt]{article}
\begin{document}

\def\slash#1{{\rlap{$#1$} \thinspace/}}

\begin{flushright} 
KEK-TH-821 \\
\end{flushright} 

\vspace{0.1cm}

\begin{Large}
       \vspace{1cm}
  \begin{center}
   {Noncommutative Gauge Theory on Fuzzy Four-Sphere \\
   and Matrix Model  }      \\
  \end{center}
\end{Large}
\vspace{1cm}

\begin{center}
{\large  Yusuke Kimura }  \\ 

\vspace{0.5cm} 
{\it Theory Group, KEK, }\\
{\it Tsukuba, Ibaraki 305-0801, Japan} \\

\vspace{0.5cm} 

{kimuray@post.kek.jp}

\vspace{0.8cm} 


\end{center}

 \vspace{1cm}

\begin{abstract}
\noindent
\end{abstract}
\noindent 
\hspace{0.4cm}
We study a noncommutative gauge theory on a fuzzy four-sphere. 
The idea is to use a matrix model with a fifth-rank Chern-Simons term 
and to expand matrices around the fuzzy four-sphere 
which corresponds to a classical solution of this model. 
We need extra degrees of freedom since algebra of coordinates 
does not close on the fuzzy four-sphere. 
In such a construction, 
a fuzzy two sphere is added 
at each point on the fuzzy four-sphere 
as extra degrees of freedom. 
It is interesting that fields on the fuzzy four-sphere have 
higher spins due to the extra degrees of freedom. 
We also consider a theory around the north pole and 
take a flat space limit. 
A noncommutative gauge theory on four-dimensional plane, 
which has Heisenberg type noncommutativity, is considered.

\newpage 

\section{Introduction}

\hspace{0.4cm}
One of the recent interesting developments in string theory is the 
appreciation of noncommutative geometry. 
The first paper which points out a relation 
between string theory and noncommutative geometry 
is \cite{wittenopensft}. 
In the paper, string field theory was formulated in terms 
of noncommutative geometry. 
Some studies of D-branes 
show further relations between string theory and 
noncommutative geometry.  
A system of $N$ coincident D-branes 
is described by the $U(N)$ Yang Mills theory. 
In this theory, $U(N)$ adjoint scalars represent 
the transverse coordinates of this system. 
Since they are given by $U(N)$ matrices, this fact suggests that 
the spacetime probed by D-branes may be 
related to noncommutative geometry. 
Noncommutative geometry also appears 
within the framework of toroidal compactification 
of matrix model \cite{CDS}. 
It is discussed in \cite{SW} that 
the world volume theory on D-branes with NS-NS two form  
background is described by noncommutative gauge theory. 
These studies suggest that noncommutative geometry 
may play a fundamental role in string theory. 

Matrix models are obtained by the dimensional 
reduction from Yang-Mills theory and 
concrete models \cite{BFSS,IKKT} are proposed to study 
M theory and string theory. 
IIB Matrix Model \cite{IKKT} is one of these proposals.
The action of this model is given by the dimensional reduction 
of ten-dimensional $\cal{N}$=1 $U(N)$ super Yang-Mills theory 
to a point; 
\begin{equation}
 S= -\frac{1}{g^{2}} Tr \left( \frac{1}{4} \left[ A_{\mu} ,A_{\nu}\right] 
         \left[A^{\mu} ,A^{\nu} \right]  
  +\frac{1}{2}\bar{\psi } \Gamma^{\mu} \left[ A_{\mu},\psi \right] \right) , 
  \label{actionIIB}
\end{equation}
where $\psi$ is a ten dimensional Majorana-Weyl spinor field, 
and $A_{\mu}$ and $\psi$ are $N \times N$ hermitian matrices. 
Indices $\mu,\nu$ run over $1$ to $10$ and they are contracted by 
Euclidean metric $\delta_{\mu\nu}$. 
This model is expected to give the constructive definition of 
type IIB superstring theory \cite{AIKKTT}. 

In the matrix model, eigenvalues of 
bosonic variables are interpreted as spacetime 
coordinates, and matter and even spacetime may 
dynamically emerge out of matrices \cite{AIKKTT,AIKKT}. 
Spacetime coordinates are represented  
by matrices and therefore noncommutative geometry 
is expected to appear. 
The idea of the noncommutative geometry 
is to modify the microscopic structure of the spacetime. 
This modification is implemented by replacing fields 
on the spacetime by matrices. 
A flat noncommutative background appears 
as a classical solution of the matrix model (\ref{actionIIB}); 
\begin{equation} 
[\hat{x}_{\mu},\hat{x}_{\nu}]=-iC_{\mu\nu}{\bf 1}, 
\label{heisenberg} 
\end{equation} 
where $C_{\mu\nu}$ is a constant second rank tensor. 
This solution preserves a part of supersymmetry. 
It was shown \cite{Li,AIIKKT} that 
noncommutative Yang-Mills theories in the flat background  
are obtained by 
expanding the matrix model around 
the flat noncommutative background; 
\begin{equation}
A_{\mu}=\hat{x}_{\mu}+\hat{a}_{\mu}(\hat{x}). 
\label{expansion}
\end{equation}
Fields on the background appear as 
fluctuations around the background, 
this implying 
the unification of spacetime and fields. 

It is important to study curved backgrounds 
since the matrix model is expected to be the theory of gravity. 
Especially how general covariance is embedded in the 
matrix model is a interesting problem. 
Some attempts to search general covariance in the matrix model 
are reported in \cite{IK,AIKO,azumakawai}. 
The IIB matrix model has only flat noncommutative  
backgrounds as classical solutions. 
In \cite{IKTW}, 
we have considered 
a three dimensional supersymmetric matrix model with 
a third-rank Chern-Simons term.  
This model has a fuzzy two-sphere as a classical solution. 
A Fuzzy two-sphere\footnote{
There are many papers about fuzzy two-sphere. 
See, for example, \cite{madore}.} 
is obtained by introducing the coordinates 
satisfying the following relations, 
\begin{equation}
[\hat{x}_{i},\hat{x}_{j}]=i\alpha\epsilon^{ijk}\hat{x}_{k}, 
\hspace{0.4cm}\hat{x}_{i}\hat{x}_{i}=\rho^{2}, 
\label{fuzzytwosphere}
\end{equation}
where $\alpha$ is a dimensionful constant. 
This algebra respects $SO(3)$ symmetry. 
These matrices are constructed from the $SU(2)$ algebra. 
The second condition is automatically satisfied 
by the quadratic Casimir. 
We showed in \cite{IKTW} that expanding the model around 
the fuzzy two-sphere solution leads to a noncommutative 
gauge theory on the fuzzy sphere. 
In \cite{yk2}, a four dimensional 
bosonic matrix model with a mass term  
was considered. 
This model has two classical solutions, a fuzzy two-sphere and 
a fuzzy two-torus. 
Using this model, we analyzed noncommutative 
gauge theories on a fuzzy sphere and a fuzzy torus. 
A fuzzy torus is obtained by introducing two 
unitary matrices 
satisfying the following relation, 
\begin{equation}
UV=e^{i\theta}VU.  
\end{equation}
It is natural to use the unitary matrices since 
the eigenvalues of them are distributed over circles. 
A fuzzy two-sphere and a fuzzy two-torus are constructed 
from finite dimensional matrices since the size of matrices 
represents the number of quanta on the noncommutative manifolds. 
On the other hand, a noncommutative plane (\ref{heisenberg}) 
cannot be constructed 
by finite size matrices since 
the extension of the plane is infinite. 
It is desirable that a classical solution 
is described by a finite size matrix since $N$ is considered as 
a cutoff parameter in the matrix model. 

As explained in the previous paragraph, 
$N$ plays the role of cutoff parameter in the case of the compact manifold. 
Let us consider fuzzy spheres as examples. 
When we introduce a cutoff parameter $N-1$ for angular momentum 
in a two-sphere, 
the number of independent functions is 
$\sum_{l=0}^{N-1}(_{3}H_{l}-_{3}H_{l-2})
=\sum_{l=0}^{N-1}(2l+1)=N^{2}$. 
Then we can replace the functions with $N \times N$ matrices, 
and algebras on the sphere become noncommutative.  
A generalization to a higher dimensional sphere is, 
however, not straightforward. 
Let us next consider a four-dimensional sphere. 
When we introduce a cutoff parameter $n$ for angular momentum, 
the number of independent functions is 
$\sum_{l=0}^{n}(_{5}H_{l}-_{5}H_{l-2})
=(n+1)(n+2)^{2}(n+3)/12$. 
This is not a square of an integer. 
In this case, we cannot construct a map from functions to matrices. 
We can restate this difficulty from the algebraic point of view. 
Algebras of a fuzzy four-sphere are constructed in \cite{castelino}, 
and the difference from the fuzzy two-sphere case is that 
the commutators of the coordinates do not close 
in the fuzzy four-sphere case. 
This fact makes the analyses of field theories on the fuzzy 
four-sphere difficult. 

Recently there are some developments in this fields. 
In \cite{horamgooram}, the authors showed that 
the matrix description of a fuzzy four-sphere 
is given by $SO(5)/U(2)$ coset 
and a fuzzy two-sphere 
appears as a fiber on a four-sphere. 
The stabilizer group of this four sphere 
is not $SO(4)$ but $U(2)$. 
The authors in \cite{zhanghu} considered the quantum Hall effect 
on a four dimensional sphere. 
It is well known that 
noncommutative geometry is naturally realized by the guiding 
center coordinates of the two-dimensional system of electrons 
in a constant magnetic field. 
Their system is considered as the generalizaion of the 
two-dimensional system, and is 
composed of particles moving in a four-dimensional 
sphere under the $SU(2)$ gauge field. 
The existence of Yang's $SU(2)$ monopole \cite{Yang} 
in the system makes the coordinates of particles 
moving in the four-dimensional space noncommutative. 
They showed that the configuration space of this system is 
locally $S^{4}\times S^{2}$. 
There are further analyses in \cite{fabinger,chenhou,karanair} 
following these papers. 

In this paper, we consider noncommutative gauge theory on 
a fuzzy four-sphere using a matrix model 
with a fifth-rank Chern-Simons term 
following these recent developments. 
In section two, we explain a matrix description of 
a fuzzy four-sphere based on \cite{castelino}. 
In section three, we consider a noncommutative 
gauge theory on fuzzy four-sphere using the matrix model. 
We expand the matrices around a classical solution of fuzzy four-sphere 
by the same way as (\ref{expansion}). 
It is shown that the Hamiltonian of the quantum Hall system 
on the four-dimensional sphere appears from this 
matrix model. 
In section four, we consider a noncommutative gauge theory 
on a noncommutative four-dimensional plane 
by taking a large $N$ limit. 
We use a technique which is similar to 
In\"{o}n\"{u}-Wigner contraction. 
Section five is devoted to summary. 
We explain our convention for gamma matrices in Appendix A. 
In Appendix B, a matrix model with a mass term is 
considered. 

\vspace{0.4cm}
\noindent 
{\bf Notations} 

Indices $\mu$, $\nu$, $\ldots$ 
and $a$, $b$, $\ldots$ run over $1$ to $5$ 
and $1$ to $4$ respectively. 
Indices $i$,$j$,$k$ run over $1$ to $3$ and 
they are used to parameterize 
internal two-dimensional spheres. 


\section{Fuzzy four-sphere construction}
\hspace{0.4cm}
A fuzzy four-sphere is considered in 
\cite{GroLP,castelino,CMT}. 
In this paper, we use the construction of \cite{castelino} 
and briefly review it here. 
The fuzzy four-sphere is constructed to satisfy the 
following two conditions, 
\begin{equation}
\epsilon^{\mu\nu\lambda\rho\sigma}
\hat{x}_{\mu}\hat{x}_{\nu}\hat{x}_{\lambda}\hat{x}_{\rho}
=C\hat{x}_{\sigma}, 
\label{foursphererelation}
\end{equation}
and
\begin{equation}
\hat{x}_{\mu}\hat{x}_{\mu}=\rho^{2}, 
\label{spherecondition}
\end{equation}
where $\rho$ is a radius of the sphere. 
This sphere respects $SO(5)$ invariance. 
Let us define matrices $\hat{G}_{\mu}$ as follows, 
\begin{equation}
\hat{x}_{\mu}=\alpha\hat{G}_{\mu},   
\end{equation}
where $\alpha$ is a dimensionful constant. 
These matrices are constructed from 
the $n$-fold symmetric tensor product of 
the five dimensional Gamma matrices
\footnote{Our notation is summarized in Appendix A.}, 
\begin{equation}
\hat{G}_{\mu}^{(n)}=\left(
\Gamma_{\mu}\otimes 1\otimes \cdots \otimes 1
+1\otimes \Gamma_{\mu} \otimes \cdots \otimes 1
+\cdots 
+1\otimes 1\otimes \cdots \otimes \Gamma_{\mu}
\right)_{Sym},
\label{defGmatrix}
\end{equation}
where Sym means that we are 
considering the completely symmetrized tensor product. 
The dimension of this $n$-fold symmetrized tensor product 
space, that is the size of these matrices, is calculated as 
\begin{equation}
N=_{4}H_{n}=_{n+3}C_{n}=\frac{1}{6}(n+1)(n+2)(n+3). 
\label{size}
\end{equation}
If we replace $\Gamma_{\mu}$ with Pauli matrices, 
$\hat{G}_{\mu}$ becomes the coordinates 
of a fuzzy two-sphere, that is 
($n+1$)-dimensional representation of $SU(2)$. 
After some calculations, we find that 
these matrices satisfy the following relations, 
\begin{equation}
\hat{G}_{\mu}^{(n)}\hat{G}_{\mu}^{(n)}=n(n+4) \equiv c, 
\label{ggc}
\end{equation}
and 
\begin{equation}
\epsilon^{\mu\nu\lambda\rho\sigma}
\hat{G}_{\mu}^{(n)}\hat{G}_{\nu}^{(n)}
\hat{G}_{\lambda}^{(n)}\hat{G}_{\rho}^{(n)}
=\epsilon^{\mu\nu\lambda\rho\sigma}
\hat{G}_{\mu\nu}^{(n)}
\hat{G}_{\lambda\rho}^{(n)}
=(8n+16)\hat{G}_{\sigma}^{(n)}, 
\label{commutationrelation} 
\end{equation}
where  
\begin{equation}
\hat{G}_{\mu\nu}^{(n)} \equiv \frac{1}{2}
[\hat{G}_{\mu}^{(n)},\hat{G}_{\nu}^{(n)}]. 
\label{GGcomm}
\end{equation}
(\ref{commutationrelation}) is also rewritten as 
\begin{equation}
\hat{G}_{\mu\nu}^{(n)}=-\frac{1}{2(n+2)}
\epsilon^{\mu\nu\lambda\rho\sigma}\hat{G}_{\lambda\rho}^{(n)}
\hat{G}_{\sigma}^{(n)}
=-\frac{1}{2(n+2)}\epsilon^{\mu\nu\lambda\rho\sigma}
\hat{G}_{\lambda}^{(n)}
\hat{G}_{\rho}^{(n)}\hat{G}_{\sigma}^{(n)}. 
\end{equation}
If we take $C$ as 
\begin{equation}
C=(8n+16)\alpha^{3}, 
\end{equation}
(\ref{foursphererelation}) is satisfied. From 
(\ref{spherecondition}) and (\ref{ggc}), 
$\rho$ and $\alpha$ are related by 
\begin{equation}
\rho^{2}=\alpha^{2}n(n+4)\equiv \alpha^{2} c. 
\end{equation}
We also have the following relations 
\begin{equation}
\hat{G}_{\mu\nu}^{(n)}\hat{G}_{\nu}^{(n)}
=4\hat{G}_{\mu}^{(n)}, 
\label{relationggg}
\end{equation}
\begin{equation}
\hat{G}_{\mu\nu}^{(n)}\hat{G}_{\nu\mu}^{(n)}=4n(n+4)=4c, 
\label{GijGji}
\end{equation} 
and
\begin{equation}
\hat{G}_{\mu\nu}^{(n)}\hat{G}_{\nu\lambda}^{(n)}
=c\delta_{\mu\lambda}
+\hat{G}_{\mu}^{(n)}\hat{G}_{\lambda}^{(n)}
-2\hat{G}_{\lambda}^{(n)}\hat{G}_{\mu}^{(n)}. 
\label{ggcgg}
\end{equation}
Commutation relations of these matrices are given by 
\begin{eqnarray} 
[\hat{G}_{\mu}^{(n)},\hat{G}_{\nu\lambda}^{(n)}]&=&2\left(
\delta_{\mu\nu}\hat{G}_{\lambda}^{(n)}-\delta_{\mu\lambda}
\hat{G}_{\nu}^{(n)}
\right), 
\label{commurelation}
\end{eqnarray}
\begin{equation}
[\hat{G}_{\mu\nu}^{(n)},\hat{G}_{\lambda\rho}^{(n)}]=2\left(
\delta_{\nu\lambda}\hat{G}_{\mu\rho}^{(n)}
+\delta_{\mu\rho}\hat{G}_{\nu\lambda}^{(n)}
-\delta_{\mu\lambda}\hat{G}_{\nu\rho}^{(n)}
-\delta_{\nu\rho}\hat{G}_{\mu\lambda}^{(n)}
\right). 
\label{commurelation2}
\end{equation}
This is the $SO(5)$ algebra. 

Here we comment on the physical meaning of $N$ and $n$ 
in brane interpretations. 
The authors of \cite{castelino} 
discussed this fuzzy four-sphere as 
a longitudinal five-brane in the context of 
BFSS matrix model. 
They showed that 
$N$ and $n$ represents the number of D-particles and 
longitudinal five-branes respectively. 
We are considering $n$ overlapping longitudinal five-branes.

The area occupied by the unit quantum 
on {\it the fuzzy four-sphere} is 
\begin{equation}
\hbar=\frac{\frac{8}{3}\pi^{2}\rho^{4} n}{N}
=\frac{16\pi^{2}\rho^{4}n}{(n+1)(n+2)(n+3)}, 
\end{equation}
where $8\pi^{2}\rho^{4}/3$ is a area of the four-sphere, and 
we have $N$ quanta in this system. 
We must draw attention to the factor $n$. 
As discussed in the next section, a fuzzy two-sphere 
is attached to each point on the four-sphere \cite{horamgooram}, 
and there are $n$ quanta on the fuzzy two-sphere. 
Therefore the number of the quanta on the fuzzy four-sphere 
is $N/n$. 
This is a different feature from 
the fuzzy two-sphere (\ref{fuzzytwosphere})
(or the fuzzy plane (\ref{heisenberg})). 
There is a degree of freedom 
at each point and there are 
$N$ points  on the fuzzy two-sphere. On the other hand, 
there are $n \sim N^{\frac{1}{3}}$  
degrees of freedom at each point and 
there are $n^{2} \sim N^{\frac{2}{3}}$ points on the 
fuzzy four-sphere. 
From the viewpoint of a noncommutative field theory 
on the fuzzy four-sphere, 
it is natural to interprete $n$ as the spin degree of freedom. 
(We explain this point in the next section.) 
Therefore 
fields on the fuzzy four-sphere 
have the spin, and the rank of it can be up to $n$. 

\vspace{0.5cm} 
Since we obtained the fuzzy four-sphere geometry, 
the next step is to investigate a field theory 
on it. 
The fact that the algebra of $\hat{x}_{\mu}$ 
does not, however, close makes investigations difficult.  
Functions on a usual classical four-sphere can be expanded by the 
spherical harmonics 
\begin{equation}
a(x)=\sum_{l=0}^{\infty}\sum_{m_{i}}
a_{lm_{i}}Y_{lm_{i}}(x), 
\end{equation}
where the spherical harmonics is given by 
\begin{equation}
Y_{lm_{i}}(x)=\frac{1}{\rho^{l}}
\sum_{a}f_{a_{1},\ldots,a_{l}}^{(lm_{i})}
x^{a_{1}}\cdots x^{a_{l}}, 
\end{equation}
where $m_{i}$ denote relevant quantum numbers. 
$f_{a_{1},\ldots,a_{l}}$ is a traceless and symmetric tensor. 
The traceless condition comes form $x_{i}x_{i}=\rho^{2}$. 
Matrices corresponding to the above functions are 
\begin{equation}
\hat{a}(\hat{x})=\sum_{l=0}\sum_{m_{i}}
a_{lm_{i}}\hat{Y}_{lm_{i}}(\hat{x}), 
\end{equation}
where 
\begin{equation}
\hat{Y}_{lm_{i}}(\hat{x})=\frac{1}{\rho^{l}}
\sum_{a}f_{a_{1},\ldots,a_{l}}^{(lm_{i})}
\hat{x}^{a_{1}}\cdots \hat{x}^{a_{l}}. 
\label{classicalharmonics}
\end{equation}
Due to the relation (\ref{GGcomm}), 
algebra among the matrix spherical harmonics 
does not close. 
This problem does not appear in the fuzzy two-sphere case. 
To overcome this difficulty, we have two strategies. 
The first one is to project out $\hat{G}_{\mu\nu}$ \cite{ramgoo} 
and 
the second one is to include $\hat{G}_{\mu\nu}$ whose 
counterpart in a usual classical sphere does not exist. 
In \cite{ramgoo} a product which closes without $\hat{G}_{\mu\nu}$ 
is constructed. 
This product is, however, non-associative. 
Since the matrix algebra has associativity, we want 
to maintain associativity 
to use matrix models. 
On the other hand, 
if we include $\hat{G}_{\mu\nu}$ to maintain associativity, 
the geometry which is constructed 
from $\hat{G}_{\mu}$ and $\hat{G}_{\mu\nu}$ 
becomes the coset manifold $SO(5)/U(2)$ \cite{horamgooram}. 
This coset is not $S^{4}$ but locally 
$S^{4} \times S^{2}$. 
Throughout this paper, we call a noncommutative space which 
is given by this coset 
a fuzzy four-sphere. 
In the next section, we consider a noncommutative gauge theory on a 
fuzzy four-sphere using a matrix model .


\section{Matrix model and Noncommutative gauge theory on fuzzy four-sphere}
\label{section:NCGT}
\hspace{0.4cm}
To investigate a noncommutative gauge theory on a fuzzy four-sphere, 
we consider the following matrix model 
\begin{equation}
S=-\frac{1}{g^{2}}Tr\left(
\frac{1}{4}[A_{\mu},A_{\nu}][A_{\mu},A_{\nu}]
+\frac{\lambda}{5}\epsilon^{\mu\nu\lambda\rho\sigma}
A_{\mu}A_{\nu}A_{\lambda}A_{\rho}A_{\sigma}
\right),
\label{action}
\end{equation} 
where $\mu,\nu,\ldots,\sigma$ run over 1 to 5 and 
$\epsilon^{\mu\nu\lambda\rho\sigma}$ is the $SO(5)$ invariant tensor. 
$A_{\mu}$ are $N\times N$ hermitian matrices and 
$\lambda$ is a dimensionful constant which 
depends on $N$. 
The indices are contracted by the Euclidean metric $\delta_{\mu\nu}$. 
Our discussions are restricted only to the bosonic sector. 
This is a reduced model of Yang-Mills action with 
a fifth rank Chern-Simons term. 
The second term is also interpreted as a so called 
Myers term \cite{myers}. 
In this interpretation, this action represents 
an effective action of D(-1)-branes in a constant 
R-R four-form background. 
Here we do not make use of such an interpretation. 

This model has the $SO(5)$ symmetry and 
the following unitary symmetry; 
\begin{equation}
A_{\mu} \rightarrow UA_{\mu}U^{\dagger}. 
\label{unitarytr} 
\end{equation} 
It also has the translation symmetry; 
\begin{equation}
A_{\mu} \rightarrow A_{\mu} + c_{\mu}{\bf 1}.
\end{equation}

The equation of motion of this action is as follows, 
\begin{equation} 
[A_{\nu},[A_{\mu},A_{\nu}]] 
+\lambda\epsilon^{\mu\nu\lambda\rho\sigma}
A_{\nu}A_{\lambda}A_{\rho}A_{\sigma}=0. 
\end{equation} 
There are two classical solutions, 
firstly diagonal commuting matrices 
\begin{equation} 
A_{\mu}=diag(x_{\mu}^{(N)},\ldots,x_{\mu}^{(1)}),  
\label{diagonalsolution}
\end{equation} 
and secondly a fuzzy four-sphere
\begin{equation} 
A_{\mu}=\hat{x}_{\mu}=\alpha\hat{G}_{\mu}^{(n)}.
\label{fuzzyfourspheresolution}
\end{equation} 
$\lambda$ is determined by the condition that 
the matrix model has a classical solution of the fuzzy four-sphere.  
We easily find from (\ref{commurelation}) that 
$\lambda$ is determined as the following value, 
\begin{equation} 
\lambda=\frac{2}{\alpha(n+2)}. 
\end{equation}
We should notice that  
a system of two four-spheres is {\sl not} a classical soliton 
since $\lambda$ depends on $n$. 

Although we add the Chern-Simons term to the Yang-Mills action, 
a reduced model of the Yang-Mills action with a mass term also has 
a fuzzy four-sphere as a classical solution, 
\begin{equation}
S=-\frac{1}{g^{2}}Tr\left(
\frac{1}{4}[A_{\mu},A_{\nu}][A_{\mu},A_{\nu}]
+8\alpha^{2}A_{\mu}A_{\mu}
\right). 
\end{equation} 
In this case, 
both of a two-sphere and a two-torus are also classical solutions. 
A matrix model with a mass term is investigated in \cite{yk2}.
In this model, a system of multiple fuzzy four-spheres is a 
classical solution. 
We give some comments on this model in Appendix B. 

We compare classical values of the action for 
two classical solutions. 
The value of the action (\ref{action}) 
for (\ref{diagonalsolution}) is 
\begin{equation} 
S=0
\end{equation} 
while 
\begin{equation} 
S=\frac{4\alpha^{2}\rho^{2}N}{5g^{2}}
=\frac{4\rho^{4}}{5g^{2}}
\frac{N}{n(n+4)}
\end{equation} 
for (\ref{fuzzyfourspheresolution}). 
\footnote{
We thank T. Azuma for pointing out a mistake in the 
previous version.  
}
We conclude that 
the diagonal commuting matrix is more stable 
than the fuzzy four-sphere solution at the classical level. 
This stuation is opposite to the case of the action 
which has an additional third rank CS term. 
In this case, a fuzzy two-sphere is stable than the diagonal 
commuting matrix. Therefore it is interesting to investigate 
classical dynamics of these solutions in the present situation. 

\vspace{0.5cm} 

Let us consider noncommutative gauge theory on the fuzzy four-sphere. 
The idea is to expand the matrices around the classical solution as in 
\cite{AIIKKT,IKTW,yk2}. 
We expand the matrices as follows, 
\begin{equation}
A_{\mu}=\hat{x}_{\mu}+ \alpha \rho \hat{a}_{\mu} 
=\alpha\rho \left( \frac{1}{\rho}\hat{G}_{\mu}+ \hat{a}_{\mu} \right). 
\label{expansionmatrices}
\end{equation}
We define $\hat{w}_{\mu\nu}$ as 
\begin{equation}
\hat{w}_{\mu\nu}\equiv i\alpha G_{\mu\nu}=
\frac{i\alpha}{2}[\hat{G}_{\mu},\hat{G}_{\nu}]
\end{equation}
where $\alpha$ has dimension of length. 
$i$ is added to make $\hat{w}_{\mu\nu}$ hermitian. 
We have the following noncommutativity on the fuzzy four-sphere 
\begin{equation}
[\hat{x}_{\mu},\hat{x}_{\nu}]=-2i\alpha\hat{w}_{\mu\nu}, 
\label{coordinatecommutation}
\end{equation}
where $\hat{w}_{\mu\nu}$ satisfies 
\begin{equation}
\epsilon^{\mu\nu\lambda\rho\sigma}\hat{w}_{\mu\nu}
\hat{w}_{\lambda\rho}
=-\alpha(8n+16)\hat{x}_{\sigma}.  
\end{equation} 
We now comment on a classical sphere. 
It is obtained by a large $n$ limit with 
the fixed radius of the sphere $\rho$. In other words, it is 
the $\alpha \rightarrow 0$ limit 
with the fixed $\rho$.
From (\ref{coordinatecommutation}), 
the coordinates commute each other 
in this limit:
\begin{equation}
[\hat{x}_{\mu},\hat{x}_{\nu}]=-2i\alpha\hat{w}_{\mu\nu}
 \sim O (\alpha \rho) \rightarrow 0.
\end{equation}
The coordinates $\hat{x}_{\mu}$ and $\hat{w}_{\mu\nu}$ 
also become commuting matrices: 
\begin{equation}
[\hat{x}_{\mu},\hat{w}_{\nu\lambda}]=0, 
\end{equation}
\begin{equation}
[\hat{w}_{\mu\nu},\hat{w}_{\lambda\rho}]=0 .
\end{equation}

Fields on a sphere are expanded by the spherical harmonics. 
A Fuzzy sphere is naturally introduced by giving 
a cutoff parameter for angular momentum. 
The spherical harmonics on the 
higher dimensional fuzzy sphere is considered in 
\cite{ramgoo,chenhou}. 
The bases are classified by the $SO(5)$ representations 
and the matrices are expanded by the irreducible 
representations of $SO(5)$. 
The irreducible representation is characterized by the Young diagram. 
It is labeled by the row length $(r_{1},r_{2})$ 
in this case. \footnote{
The representation of $SO(5)$ is summarized in \cite{ramgoo}.}
Only the representations with $r_{2}=0$ correspond to 
the classical sphere.
Summing up the dimensions of all irreducible representations 
with the condition $n\geq r_{1} \geq r_{2}$ leads 
to the square of $N$. 
We write the spherical harmonics abstractly as follows, 
\begin{equation} 
\hat{a}(\hat{x},\hat{w})=\sum_{r_{1}=0}^{n}
\sum_{r_{2},\tilde{m}_{i}} 
a_{r_{1}r_{2}\tilde{m}_{i}}
\hat{Y}_{r_{1}r_{2}\tilde{m}_{i}}(\hat{x},\hat{w}), 
\end{equation} 
where $\tilde{m}_{i}$ denote relevant quantum numbers. 
It is important that we have a cutoff parameter for 
angular momentum $r_{1}$ at $n$ \cite{ramgoo}. 
If we set $w_{\mu\nu}=0$, 
$\hat{Y}_{r_{1},\tilde{m}_{i}}$ becomes 
the usual spherical harmonics (\ref{classicalharmonics}). 
We need to assume that the fields 
depend not only on $\hat{x}$ but also on $\hat{w}$. 
If we consider a function corresponding to 
the above matrix, 
\begin{equation} 
a(x,w)=\sum_{r_{1}=0}^{n}\sum_{r_{2},\tilde{m}_{i}}
a_{r_{1}r_{2}\tilde{m}_{i}}
Y_{r_{1}r_{2}\tilde{m}_{i}}(x,w), 
\end{equation} 
a product of fields becomes noncommutative 
and associative. 
We note that the noncommutativity is produced by $\hat{w}_{\mu\nu}$. 
In this construction, 
$\hat{w}_{\mu\nu}$ form fuzzy two-sphere algebras. 
This fact means that the noncommutativity 
on the fuzzy four-sphere 
is produced by the fuzzy two-sphere, as 
it will be shown later.  

When we consider a field theory corresponding to 
the matrix model around the noncommutative background, 
an adjoint action of $\hat{G}_{\mu}$ is expected to become 
the following derivative operator; 
\begin{equation}
Ad\left(\hat{G}_{\mu} \right)\rightarrow -2i\left(
w_{\mu\nu}\frac{\partial}{\partial x_{\nu}}
- x_{\nu}\frac{\partial}{\partial w_{\mu\nu}}
\right), 
\end{equation}
and adjoint action of $\hat{G}_{\mu\nu}$ becomes 
\begin{equation}
Ad\left(\hat{G}_{\mu\nu} \right)\rightarrow 2\left(
x_{\mu}\frac{\partial}{\partial x_{\nu}}
-x_{\nu}\frac{\partial}{\partial x_{\mu}}
-w_{\mu\lambda}\frac{\partial}{\partial w_{\lambda\nu}}
+w_{\nu\lambda}\frac{\partial}{\partial w_{\lambda\mu}}
\right), 
\label{adjointGmunu}
\end{equation}
where derivative of $w_{\mu\nu}$ is defined as 
\begin{equation}
\frac{\partial w_{\lambda\rho}}{\partial w_{\mu\nu}}
=\delta_{\mu\lambda}\delta_{\nu\rho}-\delta_{\mu\rho}\delta_{\nu\lambda}. 
\end{equation}
The first two terms in (\ref{adjointGmunu}) correspond to 
orbital parts and the last two terms correspond to {\it isospin} parts. 
We give a physical interpretation of these operators in 
(\ref{eigebvalueofadG}).

We next show that a fuzzy two-sphere is attached to each point 
on the fuzzy four-sphere \cite{horamgooram}, 
and it leads to the spin of fields. 
We can always diagonalize a matrix $\hat{G}_{\mu}$ 
out of the five matrices. 
We diagonalize $\hat{x}_{5}=\alpha\hat{G}_{5}$ 
as in Appendix A. 
We can construct the $SU(2)\times SU(2)$ algebra from 
the $SO(4)$ algebra 
which is a sub-algebra of the $SO(5)$ algebra; 
\begin{equation}
[\hat{N}_{i},\hat{N}_{j}]=i\epsilon_{ijk}\hat{N}_{k}, 
\end{equation} 
\begin{equation}
[\hat{M}_{i},\hat{M}_{j}]=i\epsilon_{ijk}\hat{M}_{k}, 
\end{equation}
\begin{equation}
[\hat{M}_{i},\hat{N}_{j}]=0,  
\end{equation}
where 
\begin{eqnarray}
\hat{N}_{1}&=&-\frac{i}{4}\left(\hat{G}_{23}-\hat{G}_{14}\right), 
\hspace{1.8cm}
\hat{M}_{1}=-\frac{i}{4}\left(\hat{G}_{23}+\hat{G}_{14}\right), \cr 
\hat{N}_{2}&=&\frac{i}{4}\left(\hat{G}_{13}+\hat{G}_{24}\right),  
\hspace{2.1cm}
\hat{M}_{2}=\frac{i}{4}\left(\hat{G}_{13}-\hat{G}_{24}\right), \cr 
\hat{N}_{3}&=&-\frac{i}{4}\left(\hat{G}_{12}-\hat{G}_{34}\right), 
\hspace{1.8cm}
\hat{M}_{3}=-\frac{i}{4}\left(\hat{G}_{12}+\hat{G}_{34}\right) .
\end{eqnarray}
$G_{ab}$ is written as 
\begin{eqnarray}
&&\hat{G}_{23}=2i (\hat{N}_{1}+\hat{M}_{1}),
\hspace{2cm}
\hat{G}_{14}=-2i(\hat{N}_{1}-\hat{M}_{1}), \cr
&&\hat{G}_{13}= -2i(\hat{N}_{2}+\hat{M}_{2}),
\hspace{1.7cm}
\hat{G}_{24} =-2i(\hat{N}_{2}-\hat{M}_{2}),\cr
&&\hat{G}_{12} =2i(\hat{N}_{3}+\hat{M}_{3}),
\hspace{2cm}
\hat{G}_{34}=-2i(\hat{N}_{3}-\hat{M}_{3}). 
\end{eqnarray}
The Casimir of each $SU(2)$ algebra is calculated from 
(\ref{commutationrelation}), 
(\ref{GijGji}) and (\ref{ggcgg}) as follows, 
\begin{eqnarray}
\hat{N}_{i}\hat{N}_{i}
&=&\frac{1}{16}\left(
c+(2n+4)G_{5}+G_{5}^{2}\right) \cr
&=&\frac{1}{16}\left(
n+G_{5}\right)\left( n+4+G_{5}\right), 
\end{eqnarray}
and 
\begin{eqnarray}
\hat{M}_{i}\hat{M}_{i}
&=&\frac{1}{16}\left(
c-(2n+4)G_{5}+G_{5}^{2}\right) \cr
&=&\frac{1}{16}\left(
n-G_{5}\right)\left( n+4-G_{5}\right). 
\end{eqnarray}
Matrices 
$\hat{M}_{i}$ and $\hat{N}_{i}$ are realized by 
$(n+G_{5}+2)/2$ and $(n-G_{5}+2)/2$ dimensional 
representation
\footnote{$tr\hat{N}_{i}$ and $tr\hat{M}_{i}$ represent 
two-brane charge in string theory interpretation. 
Since these are given by finite dimensional 
$SU(2)$ matrices, 
two-brane charge vanishes on each point on four-sphere. 
} of $SU(2)$ 
respectively. 
The point $\hat{G}_{5}=G_{5}$ consists of 
$(n+G_{5}+2)\cdot (n-G_{5}+2)/4$ eigenvalues. 
If we sum up the contributions 
from $G_{5}=n,n-2,\ldots,-n+2,-n$ as 
\begin{equation}
\sum_{G_{5}=-n}^{G_{5}=n}\left(\frac{n+G_{5}+2}{2} \right)
\cdot \left(\frac{n-G_{5}+2}{2}\right)  
=\frac{1}{6}(n+1)(n+2)(n+3), 
\end{equation}
we obtain the size $N$ of the matrix .  

At the north pole, the Casimirs of $\hat{N}_{i}$ and $\hat{M}_{i}$ 
are given by 
\begin{eqnarray}
\hat{N}_{i}\hat{N}_{i}
=\frac{n(n+2)}{4}
\label{Ncasimirnorth}
\end{eqnarray}
and 
\begin{eqnarray}
\hat{M}_{i}\hat{M}_{i}
&=&0 
\label{Mcasimirnorth}
\end{eqnarray}
respectively. 
Then we have a fuzzy two-sphere, 
which is given by the ($n+1$)-dimensional representation 
of $SU(2)$, at the north pole. 
The radius of the two-sphere is given by 
$\sigma^{2}=\alpha^{2}n(n+4)/4$ and it
is comparable with that of the four-sphere, 
which is given by $\rho^{2}=\alpha^{2}n(n+4)$. 
Since the fuzzy four-sphere has $SO(5)$ symmetry, 
we can state that a fuzzy two-sphere, which is given by 
the ($n+1$)-dimensional representation of $SU(2)$, 
is attached to each point on the fuzzy four-sphere. 
We can regard this two-sphere as the internal two-dimensional space. 
Fields have quantum numbers corresponding to the $SU(2)$ angular 
momentum. 
We next show that these extra degrees of freedom can be 
interpreted as spins.  

Generators of Lorentz transformation are given by 
$\hat{G}_{ab}$. Fields are transformed under the Lorentz 
transformation as follows 
\begin{eqnarray}
&&e^{i\hat{G}_{ab}\omega_{ab}}\hat{a}(\hat{x},\hat{w})
e^{-i\hat{G}_{ab}\omega_{ab}} \cr 
&=&\hat{a}(\hat{x},\hat{w})
+i\omega_{ab}Ad(G)_{ab}\hat{a}(\hat{x},\hat{w}) \cr 
&\rightarrow &a(x,w) 
+2i\omega_{ab}\left(
x_{a}\frac{\partial}{\partial x_{b}}
-x_{b}\frac{\partial}{\partial x_{a}}
-w_{ac}\frac{\partial}{\partial w_{cb}}
+w_{bc}\frac{\partial}{\partial w_{ca}}
\right)a(x,w)  \cr 
&=&
a(x,w) 
+2i\omega_{ab}\left(
x_{a}\frac{\partial}{\partial x_{b}}
-x_{b}\frac{\partial}{\partial x_{a}} \right)
a(x,w)
-4i(\theta_{i}+\omega_{i})
\epsilon_{ijk}N_{j}\frac{\partial}{\partial N_{k}}
a(x,w)
\label{lorentztrans}
\end{eqnarray}
where $\theta_{i}=(\omega_{23},\omega_{31},\omega_{12})$ 
and $\omega_{i}=(\omega_{41},\omega_{42},\omega_{43})$.
Transformation from the third line to the fourth line 
is done at the north pole. It must be noted that 
such a rewriting is valid on each point on the sphere. 
The second term in the last equation 
shows the angular momentum part. 
The third term shows that the fields 
have spin angular momentum. 
The point is that the $SU(2)$ spin takes 
only the integer values, 
$0$, $1$, $\cdots$, $n-1$, $n$. 
The rank of the spin is finite 
since the sizes of $SU(2)$ matrices $\hat{N}_{i}$ 
are $n+1$. 

When we do Taylor expansion, a field with respect to 
coordinates $N_{i}$, 
\begin{eqnarray}
a(x,w)
&=&a(x,0)
+\left.N_{i_{1}}
\frac{\partial a(x,N)}
{\partial N_{i_{1}} }\right|_{N=0}
+\cdots +\left.
\frac{1}{n!} 
N_{i_{1}}N_{i_{2}}\cdots N_{i_{n}} 
\frac{\partial^{n} a(x,N)} 
{\partial N_{i_{1}}\partial N_{i_{2}} 
\cdots \partial N_{i_{n}} }\right|_{N=0} \cr
&\equiv& a(x)
+N_{i_{1}}\tilde{a}_{i_{1}}(x)
+\cdots +
\frac{1}{n!} 
N_{i_{1}}N_{i_{2}}\cdots N_{i_{n}} 
\tilde{a}_{i_{1},i_{2},\ldots,i_{n}}(x) 
\label{taylorexpansion}
\end{eqnarray}
the first term is a scalar field and the $m+1$-th term 
represents a spin $m$ field. 
This expansion is done at the north pole. 
Because the noncommutativity is produced 
by the fuzzy two-sphere, 
the product between the scalar field has the spin 
degrees of freedom.  
If we remove the fuzzy two-sphere 
from the four-sphere, the product 
becomes commutative. 
Such a product is considered in \cite{ramgoo} and 
it is commutative and {\it non-associative}. 

We now consider an action of 
noncommutative gauge theory on the fuzzy four-sphere. 
It is obtained form the matrix model action (\ref{action}) 
by expanding matrices around the classical solution
corresponding to the fuzzy four-sphere as in 
(\ref{expansionmatrices}); 
\begin{eqnarray}
S&=&-\frac{(\alpha\rho)^{4}}{g^{2}}Tr\left(
\frac{1}{4}\hat{F}_{\mu\nu}\hat{F}_{\mu\nu}
-\frac{9\lambda}{40(\alpha\rho)^{2}}
\epsilon^{\mu\nu\lambda\rho\sigma}
[A_{\mu},A_{\nu}][A_{\lambda},A_{\rho}]A_{\sigma} \right. \cr 
&&\left. \hspace{3cm}
-\frac{\lambda^{2}}{16(\alpha\rho)^{2}}
f^{\mu\nu\lambda\rho\sigma\tau}
[A_{\mu},A_{\nu}]A_{\lambda}[A_{\rho},A_{\sigma}]A_{\tau}
\right), 
\label{actionfoursphere}
\end{eqnarray} 
where 
\begin{eqnarray}
f^{\mu\nu\lambda\rho\sigma\tau}&=&
\epsilon^{\alpha\beta\mu\nu\lambda}
\epsilon^{\alpha\beta\rho\sigma\tau} \cr
&=&2\delta_{\mu\rho}\left(\delta_{\nu\sigma}\delta_{\lambda\tau}
-\delta_{\nu\tau}\delta_{\lambda\sigma}\right)
-2\delta_{\mu\sigma}\left(\delta_{\nu\rho}\delta_{\lambda\tau}
-\delta_{\nu\tau}\delta_{\lambda\rho}\right)
+2\delta_{\mu\tau}\left(\delta_{\nu\rho}\delta_{\lambda\sigma}
-\delta_{\lambda\rho}\delta_{\nu\sigma}\right). 
\end{eqnarray} 
$\hat{F}_{\mu\nu}$ is a gauge covariant field strength, 
which is defined by 
\begin{eqnarray}
\hat{F}_{\mu\nu}&\equiv& \frac{1}{(\alpha\rho)^{2}}
\left( [A_{\mu},A_{\nu}]
+\lambda\epsilon^{\mu\nu\lambda\rho\sigma}
A_{\lambda}A_{\rho}A_{\sigma} \right)\cr 
&=&\frac{1}{(\alpha\rho)^{2}}\left( 
[A_{\mu},A_{\nu}]+\frac{1}{2}
\lambda\epsilon^{\mu\nu\lambda\rho\sigma}
[A_{\lambda},A_{\rho}]A_{\sigma}\right) \cr 
&=&
\left[\frac{1}{\rho}\hat{G}_{\mu},\hat{a}_{\nu}\right]
-\left[\frac{1}{\rho}\hat{G}_{\nu},\hat{a}_{\mu}\right]
+[\hat{a}_{\mu},\hat{a}_{\nu}] \cr 
&&+\alpha\rho
\lambda\epsilon^{\mu\nu\lambda\rho\sigma}
\left(
\frac{1}{\rho^{2}}\hat{G}_{\lambda\rho}\hat{a}_{\sigma}
+\left[\frac{1}{\rho}\hat{G}_{\lambda},\hat{a}_{\rho}\right]
\left(\frac{1}{\rho}\hat{G}_{\sigma}+\hat{a}_{\sigma}\right)
\right). 
\label{fieldstrength}
\end{eqnarray}
Since $A_{\mu}$ is a covariant quantity
(we explain in the next paragraph), 
the gauge covariance of $\hat{F}_{\mu\nu}$ is manifest. 
The second and third terms in 
(\ref{actionfoursphere}) 
give gauge invariant interaction terms. 

The gauge symmetry in this noncommutative 
gauge theory comes from the unitary symmetry in 
the matrix model. 
For an infinitesimal transformation 
$U=\exp(i\hat{\lambda}) \sim 1+i\hat{\lambda}$ 
in (\ref{unitarytr}), 
a fluctuation around the fixed background 
transforms as 
\begin{eqnarray}
\delta \hat{a}_{\mu}(\hat{x},\hat{w}) 
=-\frac{i}{\rho}
[\hat{G}_{\mu},\hat{\lambda}(\hat{x},\hat{w})]
+i[\hat{\lambda}(\hat{x},\hat{w}),
\hat{a}_{\mu}(\hat{x},\hat{w})]. 
\label{gaugetr}
\end{eqnarray} 
The corresponding transformation in the field theory is 
\begin{eqnarray}
\delta a_{\mu}(x,w) 
=\frac{2}{\rho}
\left(
w_{\mu\nu}\frac{\partial}{\partial x_{\nu}}
- x_{\nu}\frac{\partial}{\partial w_{\mu\nu}}
\right) 
\lambda(x,w)
+i[\lambda(x,w),
a_{\mu}(x,w)]_{\star}. 
\end{eqnarray} 
It is known that this gauge transformation contains many 
degrees of freedom which do not exist 
in a ordinary commutative gauge theory. 
When we expand $\hat{\lambda}$ as 
\begin{equation}
\hat{\lambda}=\lambda_{0}+\epsilon^{\mu}\hat{G}_{\mu}
+\epsilon^{\mu\nu}\hat{G}_{\mu\nu}+O\left(G^{2}\right), 
\end{equation}
the contribution from the second term 
gives the rotation (\ref{lorentztrans})
and the constant shift of the field in (\ref{gaugetr}).  
(If we consider the theory around the north pole, $\hat{G}_{ab}$ and 
$\hat{G}_{5a}$ give rotation and translation respectively.)

Trace in the matrix model action (\ref{actionfoursphere}) 
corresponds to 
the integration on the coset in 
the corresponding noncommutative field theory.
It is noteworthy that this integral is taken over 
six-dimensional space, the fuzzy four-sphere and 
the internal two-dimensional space. 

We now discuss the Laplacian. 
It is natural to use $Ad(\hat{G}_{\mu\nu})^{2}$ as 
the Laplacian since it is the quadratic Casimir of $SO(5)$ 
and it is the generator of the rotation. 
We, however, have another choice $Ad(\hat{G}_{\mu})^{2}$. 
$\hat{G}_{\mu}$ and $\hat{G}_{\mu\nu}$ form $SO(5,1)$ algebra. 
Both of $Ad(\hat{G}_{\mu})^{2}$ and $Ad(\hat{G}_{\mu\nu})^{2}$ 
are the invariants of $SO(5)$. 
From the viewpoint of the matrix model, $Ad(\hat{G}_{\mu})^{2}$ 
appears naturally as the Laplacian since 
we are expanding the matrices around 
the coordinates $\hat{x}_{\mu}=\alpha\hat{G}_{\mu}$. 
After adding a gauge fixing term, the kinetic term becomes 
\begin{eqnarray} 
S_{kinetic}
&=&\frac{(\alpha\rho)^{4}}{2g^{2}}
Tr\left(\hat{a}_{\nu}\left[\frac{\hat{G}_{\mu}}{\rho},
\left[\frac{\hat{G}_{\mu}}{\rho},\hat{a}_{\nu}
\right]\right]\right) \cr 
&=&-\frac{2(\alpha\rho)^{4}}{g^{2}}Tr\left(\hat{a}_{\tau}
\left(
\frac{\partial^{2}}{\partial x_{\mu}\partial x_{\mu}} 
-\frac{x_{\mu}x_{\nu}}{\rho^{2}}
\frac{\partial^{2}}{\partial x_{\mu}\partial x_{\nu}} 
-\frac{4x_{\mu}}{\rho^{2}}
\frac{\partial}{\partial x_{\mu}}  \right. \right.\cr 
&&\left.\left. 
-\frac{2w_{\mu\nu}x_{\lambda}}{\rho^{2}} 
\frac{\partial^{2}}{\partial x_{\nu}\partial w_{\mu\lambda}} 
+\frac{x_{\nu}x_{\lambda}}{\rho^{2}} 
\frac{\partial^{2}}{\partial w_{\mu\nu}\partial w_{\mu\lambda}} 
-\frac{w_{\mu\lambda}}{\rho^{2}}
\frac{\partial}{\partial w_{\mu\lambda}} 
\right)\hat{a}_{\tau}\right). 
\label{kinetic}
\end{eqnarray} 
The first three terms correspond to the usual 
Laplacian of a four-sphere.
Let us investigate the spectrum of the kinetic term. 
It is calculated as follows, 
\begin{equation}
\frac{1}{4}[\hat{G}_{\mu},[\hat{G}_{\mu},
\hat{Y}_{r_{1}r_{2}}]]
=\left(r_{1}(r_{1}+3)-r_{2}(r_{2}+1)\right)
\hat{Y}_{r_{1}r_{2}}. 
\label{eigebvalueofadG}
\end{equation}
On the other hand, 
the spectrum of $Ad(\hat{G}_{\mu\nu})^{2}$ 
is calculated as follows, 
\begin{equation}
-\frac{1}{8}
[\hat{G}_{\mu\nu},[\hat{G}_{\mu\nu},
\hat{Y}_{r_{1}r_{2}}]]
=\left(r_{1}(r_{1}+3)+r_{2}(r_{2}+1)\right)
\hat{Y}_{r_{1}r_{2}}.
\end{equation}
The second term is regarded as a mass term from 
the four-dimensional point of view.
It is found that 
the eigenvalue of $Ad(G_{\mu})^{2}$ is identical with the eigenvalue 
of the Hamiltonian which appeared in 
the four-dimensional quantum Hall system \cite{zhanghu}. 
The Hamiltonian describes the system of 
a single particle moving on the four-dimensional sphere 
under the $SU(2)$ monopole background. 
Therefore 
$Ad(\hat{G}_{\mu})$ is interpreted as the covariant derivative 
under the $SU(2)$ monopole background.

\vspace{0.2cm}

We have so far discussed the $U(1)$ noncommutative gauge theory 
on the fuzzy sphere. 
A generalization to $U(m)$ gauge group is realized 
by the following replacement: 
\begin{equation}
\hat{x}_{\mu} \rightarrow \hat{x}_{\mu}\otimes{\bf 1}_{m}. 
\end{equation}
\noindent 
$\hat{a}$ is also replaced as follows:
\begin{equation}
\hat{a} \rightarrow 
\sum_{a=1}^{m^{2}}\hat{a}^{a}\otimes T^{a}, 
\end{equation}
\noindent 
where $T^{a} (a=1,\cdots, m^{2})$ denote the generators of $U(m)$.


\section{Noncommutative gauge theory on noncommutative four-plane}
\hspace{0.4cm}
In this section, 
we consider a noncommutative gauge theory 
on a noncommutative four-plane, which arises as a large $N$ limit 
from the fuzzy four-sphere. 
This limit corresponds to considering a subspace in this system. 
Let us consider a theory around the north pole, 
that is $\hat{x}_{5}\sim \rho$ 
($\hat{G}_{5}\sim \rho/\alpha \sim n$). 
By virtue of the $SO(5)$ symmetry, this discussion is 
without loss of generality. 

As discussed in (\ref{Ncasimirnorth}) and 
(\ref{Mcasimirnorth}), a fuzzy two-sphere, 
which is given by the ($n+1$)-dimensional representation 
of $SU(2)$, 
is attached 
to the north pole. 
The commutation relation of $\hat{G}_{a}$ 
is rewritten as 
\begin{equation}
[\hat{G}_{a},\hat{G}_{b}]=2\hat{G}_{ab}
=4i\eta_{ab}^{i}\hat{N}_{i}. 
\label{ncrelation}
\end{equation}
The t' Hooft symbol $\eta_{ab}^{i}$
\footnote{
$\eta_{ab}^{i}$ satisfies 
$\eta_{ab}^{i}\eta_{ab}^{j}=4\delta_{ij}$ and 
$\eta_{ab}^{i}\eta_{ac}^{i}=3\delta_{bc}$. 
The nonzero values are  
$\eta_{23}^{1}=\eta_{41}^{1}=\eta_{31}^{2}=
\eta_{42}^{2}=\eta_{12}^{3}=\eta_{43}^{3}=1$.
} 
is introduced here as 
\begin{equation}
\eta_{ab}^{i}=\epsilon_{iab4}-\delta_{ia}\delta_{4b}
+\delta_{ib}\delta_{4a}
\end{equation}
where $i=1,2,3$ and $a,b=1,2,3,4$. 
It is interesting that 
the noncommutativity has an internal $SU(2)$ index, 
and it is used to label the spin indices of the fields. 
This equation (\ref{ncrelation}) also appeared in \cite{zhanghu}. 
If we use the t' Hooft symbol, the derivative of $G_{ab}$ is rewritten as 
\begin{equation}
\frac{\partial}{\partial G_{ab}}=-\frac{i}{2}\eta_{ab}^{i}
\frac{\partial}{\partial N_{i}}.
\end{equation}

Commutation relations 
(\ref{commurelation}) and (\ref{commurelation2}) 
are rewritten around the north pole 
as follows, 
\begin{eqnarray}
&&[\hat{G}_{ab},\hat{G}_{c}]=2\left(
\delta_{bc}\hat{G}_{a}-\delta_{ac}\hat{G}_{b}
\right), \cr
&&[\hat{G}_{5a},\hat{G}_{b}]=2\delta_{ab}n , 
\label{wignerrelation1}
\end{eqnarray}
\begin{eqnarray}
&&[\hat{G}_{ab},\hat{G}_{cd}]=2\left(
\delta_{bc}\hat{G}_{ad}+\delta_{ad}\hat{G}_{bc}
-\delta_{ac}\hat{G}_{bd}-\delta_{bd}\hat{G}_{ac}
\right), \cr
&&[\hat{G}_{5a},\hat{G}_{5b}]=-2\hat{G}_{ab}, \cr
&&[\hat{G}_{5a},\hat{G}_{bc}]=2\left(
\delta_{ab}\hat{G}_{5c}-\delta_{ac}\hat{G}_{5b}\right).  
\label{wignerrelation2}
\end{eqnarray} 
It is natural to regard $\hat{G}_{5a}$ as 
the momentum matrices since they are canonical conjugate 
to $\hat{G}_{a}$. 
If we define 
$\hat{p}_{a}=\alpha^{-1}i \hat{G}_{5a}$, 
the commutation relations become 
\begin{eqnarray}
&&[\hat{x}_{a},\hat{x}_{b}]=2\alpha^{2}\hat{G}_{ab}, \cr
&&[\hat{G}_{ab},\hat{x}_{c}]=2\left(
\delta_{bc}\hat{x}_{a}-\delta_{ac}\hat{x}_{b}
\right), \cr
&&[\hat{p}_{a},\hat{p}_{b}]=2\alpha^{-2}\hat{G}_{ab}, \cr
&&[\hat{p}_{a},\hat{G}_{bc}]=2\left(
\delta_{ab}\hat{p}_{c}-\delta_{ac}\hat{p}_{b}\right) \cr 
&&[\hat{p}_{a},\hat{x}_{b}]=2i\delta_{ab}n. 
\end{eqnarray}
From (\ref{ggcgg}), we obtain 
\begin{equation}
\alpha^{2}\hat{p}_{a}\hat{p}_{a}
=\frac{\hat{x}_{a}\hat{x}_{a}}{\alpha^{2}}. 
\end{equation} 
Momentum space also form the fuzzy four-sphere. 
Note that 
momentum matrices and coordinate matrices are different 
on the fuzzy four-sphere. 
In the fuzzy two-sphere case, they are given by the same matrices. 
They are different in general and we can see such examples 
in \cite{HoYeh,CorSch,PMHo}. 

In order to take a large radius limit, 
we rescale matrices $\hat{G}_{a}$, $\hat{G}_{ab}$ 
and $\hat{G}_{a5}$ as 
\begin{equation}
\hat{G}_{a}^{\prime}=\frac{1}{\sqrt{n}}\hat{G}_{a}, 
\hspace{0.3cm}
\hat{G}_{ab}^{\prime}=\frac{1}{\sqrt{n}}\hat{G}_{ab}, 
\hspace{0.3cm}
\hat{G}_{a5}^{\prime}=\frac{1}{\sqrt{n}}\hat{G}_{a5}.  
\end{equation}
Then the commutation relation of $\hat{G}_{a}^{\prime}$ 
becomes 
\begin{equation}
[\hat{G}_{a}^{\prime},\hat{G}_{b}^{\prime}]=
\frac{2}{\sqrt{n}}\hat{G}_{ab}^{\prime}, 
\end{equation}
or 
\begin{eqnarray}
\epsilon^{abcd}\hat{G}_{a}^{\prime}\hat{G}_{b}^{\prime}
\hat{G}_{c}^{\prime}\hat{G}_{d}^{\prime}  
=\frac{1}{n}
\epsilon^{abcd}\hat{G}_{ab}^{\prime}\hat{G}_{cd}^{\prime}
=8. 
\label{ncplane}
\end{eqnarray}  
After the rescaling, the commutation relations become as follows, 
\begin{eqnarray}
[\hat{G}_{a}^{\prime},\hat{G}_{bc}^{\prime}]&=&\frac{2}{\sqrt{n}}
\left(\delta_{ac}\hat{G}_{b}^{\prime}
-\delta_{ab}\hat{G}_{c}^{\prime}
 \right), \cr 
[\hat{G}_{5a}^{\prime},\hat{G}_{b}^{\prime}]
&=&2\delta_{ab}, 
\end{eqnarray}
and 
\begin{eqnarray}
[\hat{G}_{ab}^{\prime},\hat{G}_{cd}^{\prime}]&=&
\frac{2}{\sqrt{n}}\left(
\delta_{bc}\hat{G}_{ad}^{\prime}+\delta_{ad}\hat{G}_{bc}^{\prime}
-\delta_{ac}\hat{G}_{bd}^{\prime}-\delta_{bd}\hat{G}_{ac}^{\prime}
\right), \cr 
[\hat{G}_{5a}^{\prime},\hat{G}_{bc}^{\prime}]&=&
\frac{2}{\sqrt{n}}\left(
\delta_{ab}\hat{G}_{5c}^{\prime}-\delta_{ac}\hat{G}_{5b}^{\prime}
\right). 
\end{eqnarray}

The radius of the four-sphere in the rescaled coordinate is 
\begin{equation}
\rho^{\prime 2}= 
\hat{x}_{i}^{\prime}\hat{x}_{i}^{\prime}
=\alpha^{2}\frac{n(n+4)}{n}=\frac{1}{n}\rho^{2}
\sim \alpha^{2}n .
\end{equation} 
Area of each quantum in the rescaled coordinates is 
\begin{equation}
\hbar=\frac{\frac{8}{3}\pi^{2}\rho^{\prime 4} n}{N}
=\frac{16\pi^{2}\rho^{\prime 4}n}{(n+1)(n+2)(n+3)}
\sim 16\pi^{2}\alpha^{4}.
\end{equation}
After rescaling, $\alpha$ represents a noncommutative scale. 
To decompactify the four-sphere, we will take $\alpha=$fixed and 
$\rho^{\prime} \rightarrow \infty$ 
(or $n\rightarrow \infty$) limit. 

From (\ref{relationggg}), we have 
\begin{equation} 
\hat{G}_{a5}\hat{G}_{5}=4\hat{G}_{a}-\hat{G}_{ab}\hat{G}_{b}. 
\label{g5ga} 
\end{equation} 
If we use this equation, $\hat{G}_{a5}$ is written in terms of 
$\hat{G}_{a}$ and $\hat{G}_{ab}$. 
Independent matrices are now $\hat{G}_{a}$ and $\hat{G}_{ab}$. 

We consider the 
adjoint actions of $\hat{G}_{a}$ and $\hat{G}_{ab}$ 
around the north pole, 
\begin{eqnarray}
Ad(\hat{G}_{a})&=&
\frac{2}{i}\left(
w_{ab}^{\prime}
\frac{\partial}{\partial x_{b}^{\prime}}
-x_{b}^{\prime}
\frac{\partial}{\partial w_{ab}^{\prime}}
\right) \cr 
&=&
\frac{2}{i}\left(w_{ab}^{\prime}
\frac{\partial}{\partial x_{b}^{\prime}}
+\frac{1}{2\alpha}x_{b}^{\prime}\eta_{ab}^{i}
\frac{\partial}{\partial N_{i}^{\prime}}
\right) 
\end{eqnarray}
and 
\begin{eqnarray}
Ad\left(\hat{G}_{ab} \right)&=&2\left(
x_{a}\frac{\partial}{\partial x_{b}}
-x_{b}\frac{\partial}{\partial x_{a}}
-w_{ac}\frac{\partial}{\partial w_{cb}}
+w_{bc}\frac{\partial}{\partial w_{ca}}
\right) \cr 
&=&2\left(
x_{a}^{\prime}\frac{\partial}{\partial x_{b}^{\prime}}
-x_{b}^{\prime}\frac{\partial}{\partial x_{a}^{\prime}}
-w_{ac}^{\prime}\frac{\partial}{\partial w_{cb}^{\prime}}
+w_{bc}^{\prime}\frac{\partial}{\partial w_{ca}^{\prime}}
\right) \cr 
&=&2\left(
x_{a}^{\prime}\frac{\partial}{\partial x_{b}^{\prime}}
-x_{b}^{\prime}\frac{\partial}{\partial x_{a}^{\prime}}
-\eta_{ac}^{i}\eta_{cb}^{j}
\left(N_{i}^{\prime}\frac{\partial}{\partial N_{j}^{\prime}}-
N_{j}^{\prime}\frac{\partial}{\partial N_{i}^{\prime}}\right)
\right)  .
\end{eqnarray}

The generators of Lorentz transformation are given by 
$\hat{G}_{ab}$ and the translation generators are given by 
$\hat{G}_{5a}$. 
In view of (\ref{g5ga}), translation generators are given by 
$\hat{G}_{a}$ and $\hat{G}_{ab}$; 
\begin{eqnarray}
Ad\left(\hat{G}_{a5}\right)&=&
\frac{1}{n}\left(4Ad\left(\hat{G}_{a}\right)
-\hat{G}_{ab} Ad\left(\hat{G}_{b}\right)
-\hat{G}_{b}Ad\left(\hat{G}_{ab}\right) \right) \cr 
&=&4\frac{\alpha^{2}}{\rho^{\prime 2}}Ad\left(\hat{G}_{a}\right)
-\frac{\alpha}{\rho^{\prime}}\hat{G}_{ab}^{\prime} Ad\left(\hat{G}_{b}\right)
-\frac{\alpha}{\rho^{\prime}}\hat{G}_{b}^{\prime}
Ad\left(\hat{G}_{ab}\right) \cr 
&=&-\hat{G}_{ab}^{\prime} Ad\left(\hat{G}_{b}\right)
+O\left( \frac{1}{\rho^{\prime}}\right).
\end{eqnarray}
If we ignore the $O(1/\rho^{\prime})$, the translation 
generator is related to the adjoint action of $\hat{G}_{a}$. 
 
Here we investigate the eigenvalues of the fuzzy two-sphere 
in the large $n$ limit.
As commented in the previous section, the noncommutativity 
on the fuzzy four-sphere is produced by the fuzzy two-sphere.  
One of the three coordinates of 
the fuzzy two-sphere $\hat{N}_{3}$ is diagonalized 
as follows 
\begin{equation}
\hat{N}_{3}=diag(n/2,n/2-1, \ldots ,-n/2+1,-n/2). 
\end{equation}
When we take the large $n$ limit, the two-brane charge 
no longer vanishes. 
It is because the contributions 
from the north pole and the south pole 
in two-sphere do decouple in this limit. 
Then $\hat{N}_{3}$ takes the value $+n/2$ or $-n/2$ 
in the large $n$ limit. 
After taking the large $n$ limit, the noncommutativity 
$G_{ab}^{\prime}$ becomes as follows 
\begin{equation}
G_{12}^{\prime}=i{\bf 1}\hspace{0.2cm}(\mbox{or}\hspace{0.2cm}
 -i{\bf 1}), 
\hspace{0.5cm} 
G_{34}^{\prime}=-i{\bf 1}\hspace{0.2cm}(\mbox{or}\hspace{0.2cm}
 i{\bf 1}),
\hspace{0.5cm}
\mbox{other components are zero}. 
\label{noncommutativityflat}
\end{equation}
Then we have obtained the six-dimensional noncommutative space, 
\begin{equation}
[G_{1}^{\prime},G_{2}^{\prime}]=i{\bf 1}, \hspace{0.3cm}
[G_{3}^{\prime},G_{4}^{\prime}]=-i{\bf 1},\hspace{0.3cm}
[N_{1}^{\prime},N_{2}^{\prime}]=\frac{i}{2}{\bf 1}
\end{equation}

We next study an action of noncommutative gauge theory 
on the noncommutative plane, which is obtained 
from the fuzzy four-sphere. 
The matrices $A_{a}$ are given by 
\begin{eqnarray}
A_{a}&=&\hat{x}_{a}+ \alpha \rho \hat{a}_{a} \cr 
&=&\alpha\hat{G}_{a}+ \alpha \rho \hat{a}_{a} \cr 
&\equiv& \alpha \rho^{\prime}D^{\prime}_{a}, 
\end{eqnarray} 
where we have rescaled the field as 
$\sqrt{n}\hat{a}_{\mu}=\hat{a}_{\mu}^{\prime}$. 
$Ad(D_{a})$ is the covariant derivative 
on the flat background. 
Gauge covariant field strength becomes 
\begin{eqnarray}
\hat{F}_{\mu\nu}&=&
[D_{\mu}^{\prime},D_{\nu}^{\prime}]
+(\alpha \rho^{\prime})\lambda
\epsilon^{\mu\nu\lambda\rho\sigma}
D_{\lambda}^{\prime}D_{\rho}^{\prime}D_{\sigma}^{\prime} \cr 
&=&
\left([D_{\mu}^{\prime},D_{\nu}^{\prime}]
+\frac{\alpha \rho^{\prime}2}{\alpha(n+2)}
\epsilon^{\mu\nu\lambda\rho\sigma}
D_{\lambda}^{\prime}D_{\rho}^{\prime}D_{\sigma}^{\prime}
\right) \cr 
&=&
\left([D_{\mu}^{\prime},D_{\nu}^{\prime}]
+\frac{\alpha^{2}}{\rho^{\prime}}
\epsilon^{\mu\nu\lambda\rho\sigma}
[D_{\lambda}^{\prime},D_{\rho}^{\prime}]D_{\sigma}^{\prime}
\right) . 
\end{eqnarray} 
The action around the north pole becomes 
\begin{eqnarray}
S&=&-\frac{(\alpha \rho^{\prime})^{4}}{g^{2}}Tr\left(
\frac{1}{4}\hat{F}_{\mu\nu}\hat{F}_{\mu\nu} 
-\frac{9}{5}\frac{\alpha \rho^{\prime}}{\alpha(n+2)}
\epsilon^{\mu\nu\lambda\rho\sigma}
D_{\mu}^{\prime}D_{\nu}^{\prime}D_{\lambda}^{\prime}
D_{\sigma}^{\prime}D_{\rho}^{\prime}
\right . \cr 
&&\left . \hspace{2cm} 
-
\frac{(\alpha \rho^{\prime})^{2}}{\alpha^{2}(n+2)^{2}}
f^{\mu\nu\lambda\rho\sigma\tau}
D_{\mu}^{\prime}D_{\nu}^{\prime}D_{\lambda}^{\prime}
D_{\rho}^{\prime}D_{\sigma}^{\prime}D_{\tau}^{\prime}
\right) \cr 
&=&-\frac{(\alpha \rho^{\prime})^{4}}{g^{2}}Tr\left(
\frac{1}{4}\hat{F}_{\mu\nu}\hat{F}_{\mu\nu}
-\frac{9}{5}
\frac{\alpha}{\rho^{\prime}}
\epsilon^{\mu\nu\lambda\rho\sigma}
[D_{\mu}^{\prime},D_{\nu}^{\prime}][D_{\lambda}^{\prime},
D_{\rho}^{\prime}]D_{\sigma}^{\prime}
\right . \cr 
&&\left . \hspace{2cm} 
-\left(\frac{\alpha}{\rho^{\prime}}\right)^{2}
f^{\mu\nu\lambda\rho\sigma\tau}
D_{\mu}^{\prime}D_{\nu}^{\prime}D_{\lambda}^{\prime}
D_{\rho}^{\prime}D_{\sigma}^{\prime}D_{\tau}^{\prime}
\right) \cr 
&=&-\frac{(\alpha \rho^{\prime})^{4}}{4g^{2}}Tr\left(
\hat{F}_{ab}\hat{F}_{ab}
+2[D_{a}^{\prime},\hat{\phi}^{\prime}]
[D_{a}^{\prime},\hat{\phi}^{\prime}]
+O\left( \frac{1}{\rho^{\prime}}\right)
\right),  
\end{eqnarray} 
where we have rewritten $\hat{a}_{5}$ as $\hat{\phi}$. 
The gauge transformation (\ref{gaugetr}) is rewritten as follows, 
\begin{equation} 
\delta a_{a}^{\prime}(x^{\prime},w^{\prime}) 
=\frac{2}{\rho^{\prime}}
\left(
w_{ab}^{\prime}\frac{\partial}
{\partial x_{b}^{\prime}}
- x_{b}^{\prime}\frac{\partial}
{\partial w_{ab}^{\prime}}
\right) 
\lambda(x^{\prime},w^{\prime})
+i[\lambda(x^{\prime},w^{\prime}),
a_{a}^{\prime}(x^{\prime},w^{\prime})]_{\star}
\end{equation}
where $G_{ab}^{\prime}=w_{ab}^{\prime}/i\alpha$ 
is given by (\ref{noncommutativityflat}). 

We now investigate the kinetic term with $O(1/\rho^{\prime})$. 
(\ref{kinetic}) becomes as follows around the north pole, 
\begin{eqnarray} 
S_{kinetic}
&=&\frac{(\alpha\rho^{\prime})^{4}}{2g^{2}}
Tr\left(\hat{a}_{b}^{\prime}\left[\frac{\hat{G}_{a}}{\rho^{\prime}},
\left[\frac{\hat{G}_{a}}{\rho^{\prime}},\hat{a}_{b}^{\prime}
\right]\right]\right) \cr 
&=&-\frac{2(\alpha\rho^{\prime})^{4}}{g^{2}}
Tr\left(\hat{a}_{b}^{\prime}
\left(\frac{w_{ab}^{\prime}w_{ac}^{\prime}}{\rho^{\prime 2}}
\frac{\partial^{2} }
{\partial x_{b}^{\prime}\partial x_{c}^{\prime}}
-\frac{4}{\rho^{\prime 2}}x_{b}^{\prime}
\frac{\partial}{\partial x_{b}^{\prime}}
\right. 
\right.\cr 
&& \left. \left.
-\frac{2}{\rho^{\prime 2}}w_{ab}^{\prime}x_{c}^{\prime}
\frac{\partial^{2}}{\partial x_{b}^{\prime}
\partial w_{ac}^{\prime}} 
-\frac{1}{\rho^{\prime 2}}w_{ab}^{\prime}
\frac{\partial}{\partial w_{ab}^{\prime}}
+\frac{1}{\rho^{\prime 2}}
x_{b}^{\prime}x_{c}^{\prime}
\frac{\partial^{2}}{\partial w_{ab}^{\prime} 
\partial w_{ac}^{\prime}}
\right) \hat{a}_{b}^{\prime} \right) \cr 
&=&-\frac{2(\alpha\rho^{\prime})^{4}}{g^{2}}
Tr\left(\hat{a}_{b}^{\prime}\left(
\frac{\partial^{2} }
{\partial x_{a}^{\prime}\partial x_{a}^{\prime}}
-\frac{x_{a}^{\prime}x_{c}^{\prime} 
-\alpha^{2}G_{a5}^{\prime}G_{c5}^{\prime}
}{\rho^{\prime 2}}
\frac{\partial^{2}}{\partial x_{a}^{\prime}
\partial x_{c}^{\prime}} 
-\frac{4}{\rho^{\prime 2}}x_{b}^{\prime}
\frac{\partial}{\partial x_{b}^{\prime}}
\right.\right. \cr 
&&\left.\left.
-\frac{2}{\rho^{\prime 2}}
\eta_{ab}^{i}\eta_{ac}^{j}
N_{i}^{\prime}x_{c}^{\prime}
\frac{\partial^{2}}
{\partial x_{b}^{\prime} \partial N_{j}^{\prime}}
-\frac{4}{\rho^{\prime 2}}N_{i}^{\prime}\frac{\partial}
{\partial N_{i}^{\prime}}
+\frac{1}{4\alpha^{2}\rho^{\prime 2}}x_{b}^{\prime}x_{b}^{\prime}
\frac{\partial^{2}}{\partial N_{i}^{\prime}\partial N_{i}^{\prime}}
\right)\hat{a}_{b}^{\prime}\right), 
\label{laplaciannorth}
\end{eqnarray} 
where we have used the following relation,  
\begin{eqnarray}
G_{ab}^{\prime}G_{bc}^{\prime}=n\delta_{ac}
-G_{a}^{\prime}G_{c}^{\prime}
-G_{a5}^{\prime}G_{5c}^{\prime} .   
\end{eqnarray}
The first three terms in (\ref{laplaciannorth})
constitute the usual Laplacian of a four-sphere.  
Only the first term and the last term 
survive in $\rho^{\prime} \rightarrow \infty$
limit. 
\footnote{The order of the first and last terms 
is $O\left(1/\rho^{\prime}\right)$ 
while that of the other terms is 
$O\left(1/\rho^{\prime 2}\right)$. 
Note that $x_{a}^{\prime}x_{a}^{\prime}
\sim O\left(\rho^{\prime}\right)$.
}
It is interesting that the gauge field and 
the scalar field propagate in the six-dimensional 
space.

\vspace{0.4cm} 

In this section, 
we have investigated a noncommutative gauge theory on a flat noncommutative 
background by taking a large radius limit of the fuzzy four-sphere 
around the north pole. 
This noncommutative plane has the Heisenberg algebra type 
noncommutativity and the symmetry of this plane 
is $SO(2)\times SO(2)$. 
Although it is desirable to have $SO(4)$ symmetry,  
it is difficult to construct a noncommutative plane 
which has higher symmetry. 
This difficulty may be related to the quantization of 
Nambu bracket \cite{Nambu}. 
If a noncommutative plane has $SO(4)$ symmetry, 
it is expected that the quantization of 
Nambu bracket is realized on it. 
Although some trials \cite{ALMY,DFST} 
are implemented, 
it is difficult to obtain consistent quantization of 
Nambu bracket.


\section{Summary and Discussions}
\hspace{0.4cm}
In this paper, we have investigated 
a noncommutative gauge theory on a fuzzy four-sphere
using a five-dimensional matrix model. 
We considered a matrix model with a fifth-rank Chern-Simons term 
since this model has a fuzzy four-sphere as a classical solution.  
By dividing matrices into backgrounds 
and fields propagating on them, 
we obtained noncommutative gauge theories on the backgrounds. 
It is worth noting that 
we expanded matrices around the coordinates of 
the fuzzy four-sphere. 
This facts supports an idea that the eigenvalues of bosonic variables 
in the matrix model represents spacetime coordinates. 
A characteristic feature of noncommutative gauge theories 
or the matrix model is that 
spacetime and fields are treated on the same footing. 

One of the difficulties to consider a noncommutative 
gauge theory on a fuzzy four-sphere is that 
algebra of the coordinates does not close. 
Because of this reason, we need extra degrees of freedom. 
By adding a fuzzy two-sphere at each point on the fuzzy four-sphere, 
we can solve this difficulty. 
These extra degrees of freedom are interpreted as spins. 
The maximum magnitude of the spin is related to the 
number of the quanta on a fuzzy two-sphere, 
which is comparable with number of the quanta on the fuzzy four-sphere. 

It is well known that 
the quantum Hall system is an example of the noncommutative 
geometry. 
As is discussed in \cite{zhanghu}, 
the quantum Hall system on the four-dimensional sphere 
is constructed by considering 
a system of particles under a $SU(2)$ gauge field. 
From the kinetic term of the noncommutative gauge 
theory action on the fuzzy four-sphere, we have 
obtained the same eigenvalue as the 
Hamiltonian of the quantum Hall system. 

The advantage of compact noncommutative manifolds is that 
one can construct them in terms of finite size 
matrices 
while a solution which represents a noncommutative plane 
cannot be constructed by finite size matrices. 
(From the viewpoint of the field theories, 
$N$ plays the role of the cutoff parameter.) 
We showed that 
a gauge theory on a noncommutative plane 
were reproduced from 
a gauge theory on 
a fuzzy four-sphere by taking a large $n$ limit.  
This noncommutative plane has the Heisenberg algebra type 
noncommutativity and the symmetry is $SO(2)\times SO(2)$. 
It is difficult to construct a more symmetric 
noncommutative plane with maintaining the associativity. 


We comment on the relation to the IIB matrix model. 
The second term in the action (\ref{action}) 
is interpreted as Myers term from the viewpoint of 
a D-brane action. 
On the other hand, 
we might expect that this five-dimensional matrix model 
is obtained from IIB matrix model 
by integrating unnecessary matrices 
since this model has the same kinds of symmetries as 
IIB matrix model. 
It may be alternatively obtained by deforming IIB matrix model. 
There may be 
a new model which includes a fuzzy four-sphere as a classical solution 
and has supersymmetry. 
Such analyses will be future problems. 

\vspace{1cm}
\begin{center}
{\bf Acknowledgments}
\end{center}
\hspace{0.4cm}
I am most grateful to S.Iso and Y.Kitazawa 
for reading the manuscript and helpful discussions. 
I also thank Y.Shibusa for valuable discussions.

\renewcommand{\theequation}{\Alph{section}.\arabic{equation}}
\appendix

\section{Notations of Gamma matrices}
\setcounter{equation}{0} 
\hspace{0.4cm}
This appendix is referred to \cite{castelino}. 
An explicit form of $4\times 4$ five-dimensional 
gamma matrices is given by 
\begin{eqnarray} 
\Gamma_{\mu}  
 &=&   \left( \begin{array}{c c}
  0 & -i\sigma_{\mu} \\ 
 i\sigma_{\mu}  & 0   \\
 \end{array} \right), \hspace{0.2cm}(\mu=1,2,3)     \cr
\Gamma_{4}  
 &=&   \left( \begin{array}{c c}
  0 & 1_{2} \\ 
  1_{2}  & 0   \\
 \end{array} \right) ,     \cr
\Gamma_{5}  
 &=&   \left( \begin{array}{c c}
  1_{2} & 0 \\ 
 0  & -1_{2}   \\
 \end{array} \right),        
\end{eqnarray}
where $\sigma_{\mu}$ is the Pauli matrices. 
They satisfy the Clifford algebra: 
\begin{equation}
\{\Gamma_{\mu} ,\Gamma_{\nu} \}=2\delta_{\mu\nu}
\hspace{0.2cm}(\mu,\nu=1,2,3,4,5).
\end{equation}

The matrices $\hat{G}_{\mu}^{(n)}$ are constructed 
as in (\ref{defGmatrix}). 
In this notation, $\hat{G}_{5}^{(n)}$ is diagonalized and 
the eigenvalues are 
\begin{equation}
\hat{G}_{5}^{(n)}=diag(n,n-2,\ldots,-n+2,-n),  
\end{equation}
where the eigenvalue $m$ has 
the degeneracy $((n+2)^{2}-m^{2})/4$. 

\section{Matrix Model with mass term}
\setcounter{equation}{0} 
\hspace{0.4cm}
Let us consider a five-dimensional matrix model 
with a mass term in this appendix while 
we considered a five-dimensional matrix model 
with a Chern-Simons term in the paper. 
We investigated a four-dimensional matrix model 
with a mass term in \cite{yk2}. 
The action is  
\begin{equation}
S=-\frac{1}{g^{2}}Tr\left(
\frac{1}{4}[A_{\mu},A_{\nu}][A_{\mu},A_{\nu}]
+8\alpha^{2}A_{\mu}A_{\mu}
\right), 
\end{equation} 
where $\mu$,$\nu$ run over $1$ to $5$. 
The indices are contracted by $\delta_{\mu\nu}$. 
This model has several classical solutions. 
The first one is a fuzzy four-sphere; 
\begin{equation}
\hat{x}_{\mu}^{S^{4}}=\alpha \hat{G}_{\mu}
\hspace{0.4cm}(\mu=1,2,3,4,5), 
\label{fourmass}
\end{equation}
the radius of the four-sphere is 
$\rho_{S^{4}}^{2}=\alpha^{2}n(n+4)$.
The second one is a fuzzy two-sphere. 
\begin{eqnarray}
\hat{x}_{\mu}^{S^{2}}&=&2\sqrt{2}\alpha \hat{L}_{\mu}
\hspace{0.4cm}(\mu=1,2,3), \cr 
&=&0
\hspace{0.6cm}(\mu=4,5), 
\label{twomass}
\end{eqnarray}
where $\hat{L}_{\mu}$ is the $N$-dimensional irreducible 
representation of $SU(2)$ , and the radius of the two-sphere is 
given by $\rho_{S^{2}}^{2}=2\alpha^{2}(N^{2}-1)$. 
A system of multiple fuzzy four-spheres is a classical solution while 
it is {\it not} in a model with a fifth rank Chern-Simons term. 
A system of a fuzzy four-sphere and a fuzzy two-sphere is also 
a classical solution;  
\begin{eqnarray} 
A_{\mu}  
 &=&   \left( \begin{array}{c c}
  \hat{x}_{\mu}^{S^{4}} & 0 \\ 
 0  &  \hat{x}_{\mu}^{S^{2}}  \\
 \end{array} \right), \hspace{0.2cm}(\mu=1,2,3)     \cr
 &=&   \left( \begin{array}{c c}
  \hat{x}_{\mu}^{S^{4}} & 0 \\ 
  0  & 0   \\
 \end{array} \right), \hspace{0.2cm}(\mu=4,5) .        
\end{eqnarray}
Fuzzy two-torus is also a classical solution, 
\begin{equation}
\hat{x}_{\mu}^{T^{2}}=\frac{2\sqrt{2}}
{\sqrt{1-\cos\left(\frac{2\pi}{N}\right)}}\hat{y}_{\mu}
\hspace{0.4cm}(\mu=1,2,3,4), 
\end{equation}
where 
\begin{equation}
UV=e^{i\frac{2\pi}{N}}VU,\hspace{0.4cm}  
U=\hat{y}_{1}+i\hat{y}_{2}, \hspace{0.2cm}
V=\hat{y}_{3}+i\hat{y}_{4}. 
\end{equation}

\vspace{0.4cm}
The values of the action for the fuzzy four-sphere (\ref{fourmass}) 
and the fuzzy two-sphere (\ref{twomass}) are 
\begin{equation} 
S_{S^{4}}=-\frac{31}{4g^{2}}n(n+4)N\alpha^{4} 
\end{equation} 
and  
\begin{equation} 
S_{S^{2}}=-\frac{8}{g^{2}}N(N^{2}-1)\alpha^{4} 
\end{equation} 
respectively. Because the first one is 
$O(-n^{6}\alpha^{4}/g^{2})$ 
and the second one is $O(-n^{9}\alpha^{4}/g^{2})$, 
the second one has a lower classical action than the first one.


\end{document}